\documentclass[sn-mathphys-num]{sn-jnl}

\usepackage{graphicx}%
\usepackage{multirow}%
\usepackage{amsmath,amssymb,amsfonts}%
\usepackage{amsthm}%
\usepackage{mathrsfs}%
\usepackage[title]{appendix}%
\usepackage{xcolor}%
\usepackage{textcomp}%
\usepackage{manyfoot}%
\usepackage{booktabs}%
\usepackage{algorithm}%
\usepackage{algorithmicx}%
\usepackage{algpseudocode}%
\usepackage{listings}%
\usepackage{varwidth}%
\usepackage{caption}%
\usepackage{tabularx}%
\usepackage{enumitem}

\theoremstyle{thmstyleone}%
\theoremstyle{thmstyletwo}%

\theoremstyle{thmstylethree}%

\begin{document}

\title[]{Harnessing Lightweight Ciphers for PDF Encryption}


\author[1]{\fnm{Aastha} \sur{Chauhan}}\email{2000520310002@ietlucknow.ac.in}

\author[1]{\fnm{Deepa} \sur{Verma}}\email{dverma.csed.cf@ietlucknow.ac.in}

\affil[1]{\orgdiv{Computer Science and Engineering}, \orgname{Insitute of Engineering and Technology}, \orgaddress{\city{Lucknow}, \country{India}}}


\abstract{Portable Document Format (PDF) is a file format which is
 used worldwide as de-facto standard for exchanging documents. In fact this document that you are currently reading has been uploaded as a PDF. Confidential information is also exchanged through PDFs. According to PDF standard ISO 3000-2:2020, PDF supports encryption to provide confidentiality of the information contained in it along with digital signatures to ensure authenticity. At present, PDF encryption only supports Advanced Encryption Standard (AES) to encrypt and decrypt information. However, Lightweight Cryptography, which is referred to as crypto for resource constrained environments has gained lot of popularity
 specially due to the NIST Lightweight Cryptography (LWC) competition announced in 2018 for which ASCON was announced as the winner in February 2023. The current work constitutes the first attempt to benchmark Java implementations of NIST LWC winner ASCON and finalist XOODYAK against the current PDF encryption standard AES. Our research reveals that ASCON emerges as a clear winner with regards to throughput when profiled using two state-of-the-art benchmarking tools YourKit and JMH.}




\maketitle

\section{Introduction}\label{sec1}

PDF was first developed by Adobe in 1992. Later in 2008 it was standardized as ISO 32000. PDF allows not only text and images to be a part of the document, but also allows animation, buttons, URL embedding, XML data embedding, digital signatures etc. The reason for the PDF to be so popular is that PDF maintains it’s file structure across different platforms. Considering the security aspect of PDF, it provides confidentiality through partial encryption. It also provides authenticity and integrity of the document through digital signatures. For digital signatures RSA and Elliptic Curve Digital Signature Algorithm (ECDSA) are used while for encryption AES is used.

Since 2008 various PDF readers have came to account. One important thing to note here is that PDF encryption and signing process occurs only once in a PDF while PDF decryption and verification of signatures are done everytime the respective PDF is opened through PDF viewer. So it is important to make these processes as light as possible so as to minimize latency while keeping the security in check. This is mainly implemented by PDF readers. Many attacks have been reported against various PDF viewers which show manipulation of PDF after signing or extraction of data from encrypted PDFs or misuse of various features that PDF provide and some loopholes which PDF readers did not take into account. These attacks were reported to the respected authorities by the authors of the papers which described them \cite{pdfcert}\cite{dangerouspaths}\cite{decryption}\cite{shadowattacks}\cite{trilliondollar}. Since then many PDF readers have updated their software which makes them resistant to these type of attacks.\

On an abstract view there is a trade off between flexibility and security of PDFs. These attacks also provide more insight to the various features that PDF provides, making them more concrete. In 2020 the PDF standard has been revised with some new features and PDF 2.0 was introduced. However for backward compatibility every PDF reader must render PDF of older versions

In this paper we have worked with PDF version 1.7 since we encounter very minimal no. of PDFs with version 2.0. Moreover we implemented our code using iText \cite{itext} library in JAVA. iText was written by Bruno Lowagie \cite{itextbook}. The source code was initially distributed as open source under the Mozilla Public License or the GNU Library General Public License open source licenses. By default iText generates PDF-1.7 as default version. With respect to encryption the only changes that were made were in the creation of O value, U value and encryption process. These factors remain constant in all PDF of same version. In our work we show the comparison among AES, ASCON and XOODYAK and how their specification being lightweight helps in encryption process being faster. So we focused majorly on PDF encryption with value of R being 4 in Encrypt dictionary.

First we start by explaining very briefly about PDF structure in Section \ref{sec3}. We then state how PDF encryption works and how it affects the PDF structure and contents. Moving onto the next part we give a brief introduction to the ciphers used. In Section 3 we show how we have replaced AES with ASCON and XOODYAK. Next we show our results in Section \ref{sec4}, comparing different parameters on each ciphers.

\section{Background}\label{sec2}

 For getting a better hold of what we have done in the later part of this paper, we need to know some basics of PDF. Among those include PDF structure, so
 as to get a better hold of Partial Encryption of PDF. We also needed to know
 about the ciphers (ASCON and XOODYAK) for better implementation of them.
 Since we have taken their implementation from github, we needed to make some
 changes i.e getting rid of any unnecessary lines of computation to suit our needs
 more accurately
 \subsection{PDF File Structure}\label{subsec2}
 PDF has a linear file structure. Any changes made on the PDF are added after the \%\%EOF bytes of the PDF. Thus any modified PDF can be reverted back manually
after changes has been made by just removing the added parts. It is divided into
4 parts mainly Header, Body, Xref Table and Trailer in that exact order. We
describe each components briefly so as to understand the encryption process of
PDF more clearly.
\subsubsection{Header}\label{subsubsec2} Header of a PDF contains \% followed by PDF-version. Next line also
consists of \% followed by 4 arbitrary characters. These characters need to be of
ASCII code 128 or greater. These two lines allows any PDF reader to interpret the file as a PDF file.
\begin{figure}[h]
    \centering
    \fbox{
        \begin{minipage}{0.9\linewidth}
        \%PDF-1.7\\
        \%âãÏÓ
        \end{minipage}
    }
     \caption{ PDF Header}
\end{figure}
 \subsubsection{Body}\label{subsubsec2} PDF body comprises of only objects. Objects which may refer to image,
 text, XML data, metadata, font description, encryption information etc. Every
 object has obj x y and endobj as delimiters where x is the object number of the
 object and y is the generation number of the object. Every object has unique
 object number which generally starts from 1. The generation number is 0 in
 almost all of the objects. Every object has attributes which is defined by a data
 structure called PDF dictionary. PDF dictionary contains a list of key value pairs
 where key is of type PDF name and value can be of any data type supported by
 PDF such as strings, hexstrings, PDF name, PDF dictionary etc.
  \subsubsection{Xref Table}\label{subsubsec2} Xref Table is part of PDF maintains a table of all the objects present in the
 PDF regardless of it being used or not. This table is maintained in increasing
 order of the object numbers. Xref Table starts with the bytes xref. The actual
 table starts from the next line. First row in this table consists of 0 followed by x number of consecutive entries in the xref table. From second row each of the x entries consists of byte offset of the current object followed by the generation
 number of the object followed by either f or n. If the current object is in use,
 the value is n otherwise f, which means the object is free and is not used in
 rendering of the PDF. The first entry in the table always refers to a null object
 with object no. 0, generation number 65535 and is free (i.e. f).

 \subsubsection{Trailer}\label{subsubsec2}Although trailer is the last part of the PDF file structure, every PDF reader
 first looks for the trailer dictionary. This is because trailer contains the reference to the root object of the PDF i.e. catalog. Trailer dictionary also contains reference to the Encrypt dictionary in case the PDF is encrypted. There exists 2 hexadecimal strings of length 16 bytes with key as ID which are generated in such a way so as to uniquely identify each PDF. Whenever a PDF is manipulated the second hexadecimal value in ID gets generated again. Following line after the trailer dictionary contains the bytes startxref. Next line consists the byte offset of Xref Table. The last line of a PDF contains bytes \%\%EOF.
\begin{figure}
    \centering
      \includegraphics[width=0.5\textwidth]{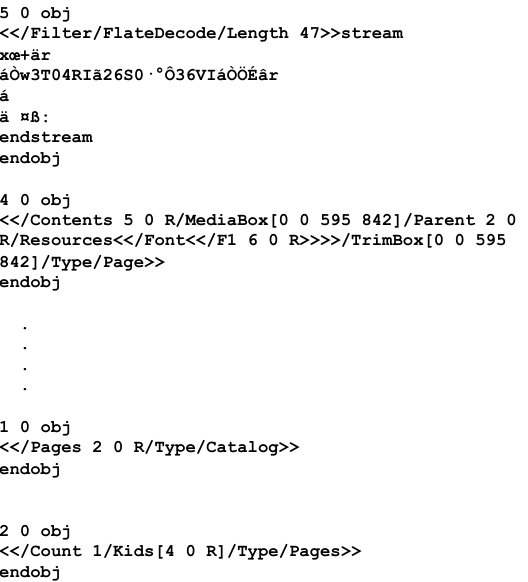}
       \caption{PDF Body}
\end{figure}
 \begin{figure}
    \centering
      \includegraphics[width=0.5\textwidth]{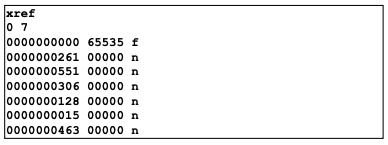}
       \caption{Xref Table}
\end{figure}

\subsection{PDF Encryption}\label{subsec2}
In earlier studies, the Data Encryption Standard (DES) was employed for securing PDF documents, demonstrating the practical application of symmetric encryption algorithms in this context. However, DES has been largely replaced by more secure algorithms like AES due to its vulnerability to brute-force attacks and relatively lower key size, which are inadequate for modern security requirements\cite{des}. PDF encryption is very flexible. PDF file as a whole is not encrypted, rather only
 the strings and streams in the PDF are encrypted. So on opening an encrypted
 PDF in a normal text editor we find that structure of the PDF remains same even
 after encryption. On opening the PDF in text editor the only way we can distinguish between an encrypted and non-encrypted PDF is the Encrypt dictionary  which will be absent in non-encrypted PDF. This is called Partial Encryption of
 PDF. PDF encrypts only strings and streams in a PDF with following exceptions

 \begin{enumerate}
     \item The hexadecimal strings in the \textbf{ID} entry of the trailer
     \item The strings present in the \textbf{Encrypt} dictionary 
     \item The signature string present in the \textbf{Contents} entry in the signature dictionary
     \item The strings present implicitly in a stream which are considered to be part
 of the stream
 \end{enumerate}
 Encryption algorithm used in PDF is either RC4 or AES. However use of RC4 has
 been deprecated since PDF-2.0 \cite{pdf2.0}. All information related to the encryption is maintained in Encrypt dictionary. Reference to the encrypt dictionary is provided through the trailer dictionary. Encrypt dictionary contains information about the algorithm used to encrypt and decrypt the document, the key length used. V value in the Encrypt dictionary depicts which algorithm is used. Length value depicts the length of the key used in bits. Length of the key must be a multiple of eight.

 Encrypted PDF consists of two kinds of password owner password and user
 password. Opening an encrypted PDF using owner password one can gain access
 to everything such as printing, copying content etc. while opening an encrypted
 \begin{figure}
  \centering
      \includegraphics[width=0.99\textwidth]{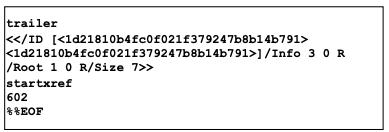}
       \caption{PDF Trailer}
\end{figure}
 PDF using user password one can gain access to only those features which are
 allowed by the owner of the document while encrypting the document. In an
 encrypted PDF there exists Encrypt dictionary which contains all the attributes of the cipher used to encrypt such as length of the key, algorithm used, cipher used to encrypt, permissions and security handler used. In Encrypt dictionary there exists P entry whose value is a 32-bit negative integer in signed two’s complement form mentioning the permissions to be given on opening the document as a set of flags using user password. From PDF-2.0 there exists a perm value which is the encrypted value of the P entry. This is only applicable if the value of R is 6. R determines the revision of the standard security handler used to interpret the Encrypt dictionary.\\

\begin{figure}
  \centering
      \includegraphics[width=0.99\textwidth]{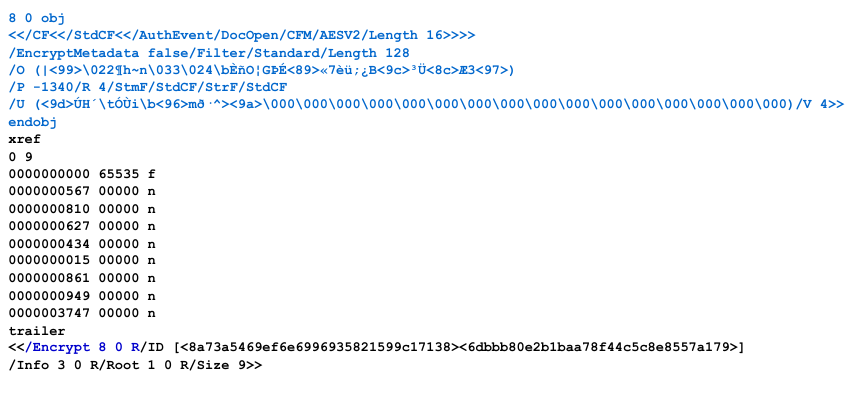}
       \caption{Example of Encrypt Dictionary in Encrypted PDF}
\end{figure}

\subsection{AES}\label{subsec2}
 Advanced Encryption Standard (AES) is a Substitution-Permutation Network
 (SPN) cipher. AES is also known as Rijndael after it’s authors Joan Daemen
 and Vincent Rijmen \cite{rijndael}\cite{rijndaelprop}\cite{sood2023literature}. AES has a state size of 128 bits or 16 words of size 1 byte each. AES supports key sizes of 128, 192, 256 bits. For each of these key variants there exists different key schedule. Since ISO 32000-2 \cite{pdf2.0}, AES-256 has
 been introduced, however we focus majorly on AES-128 since it is still majorly used in PDF encryption.\\
  \begin{figure}
  \centering
      \includegraphics[width=0.99\textwidth]{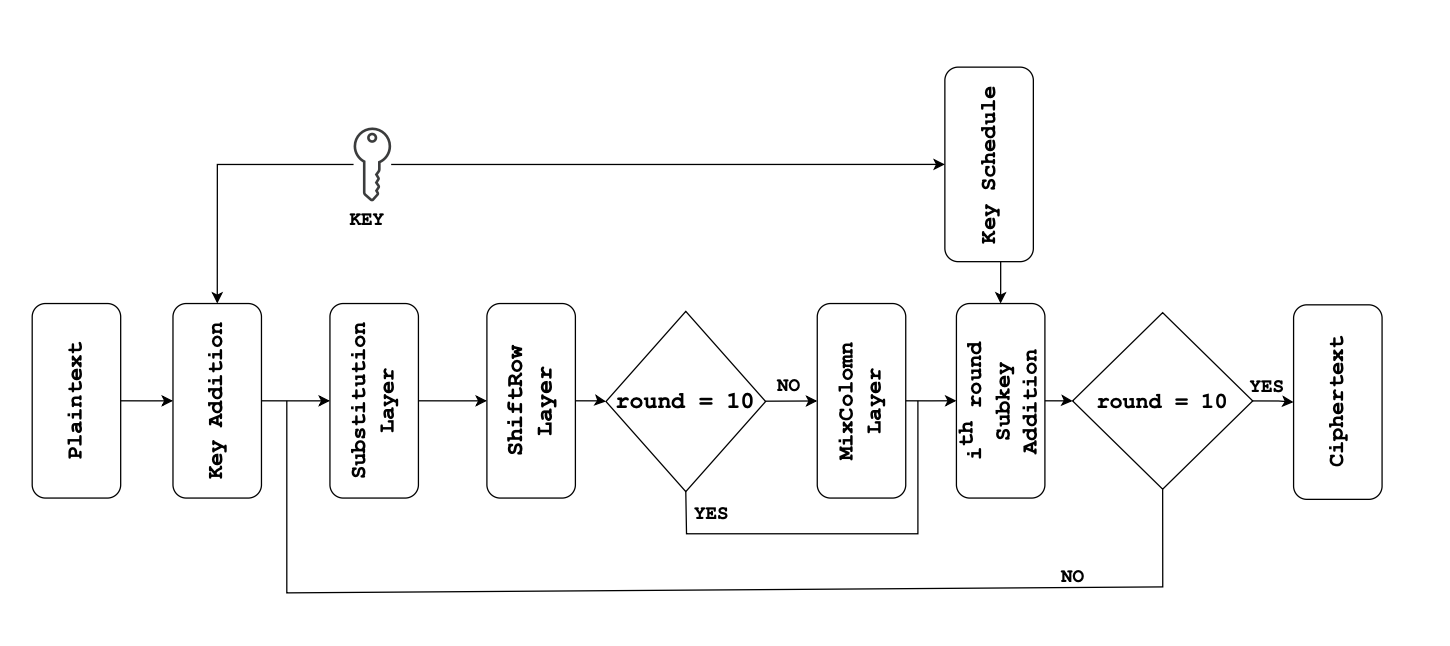}
       \caption{Encryption Process of AES}
\end{figure}
\subsection{ASCON}\label{subsec2}
 The CAESAR competition \cite{caesar} was a cryptographic competition that was active from 2013 to 2017, where numerous cipher were introduced for suitable use
 case. ASCON is a submission by Dobraunig et al \cite{ascon}\cite{asconv1.2}. ASCON was one of the five finalist of the competition and was suitable for light-weight application use case. Recently ASCON has been announced as LWCstandard by NIST \cite{nistlwc}.
 ASCON is based on a sponge-like construction with a state size of 320 bits
 (consisting of five 64-bit words x0,....,
 x4). ASCON has mainly two versions,
 ASCON-128 and ASCON-96, with different security level. In this paper we will
 focus on ASCON-128. ASCON has Sponge like structure. ASCON is divided into
 4 divisions, initialization, processing associated data, processing the plaintext,
 and finalization.
   \begin{figure}
  \centering
      \includegraphics[width=0.99\textwidth]{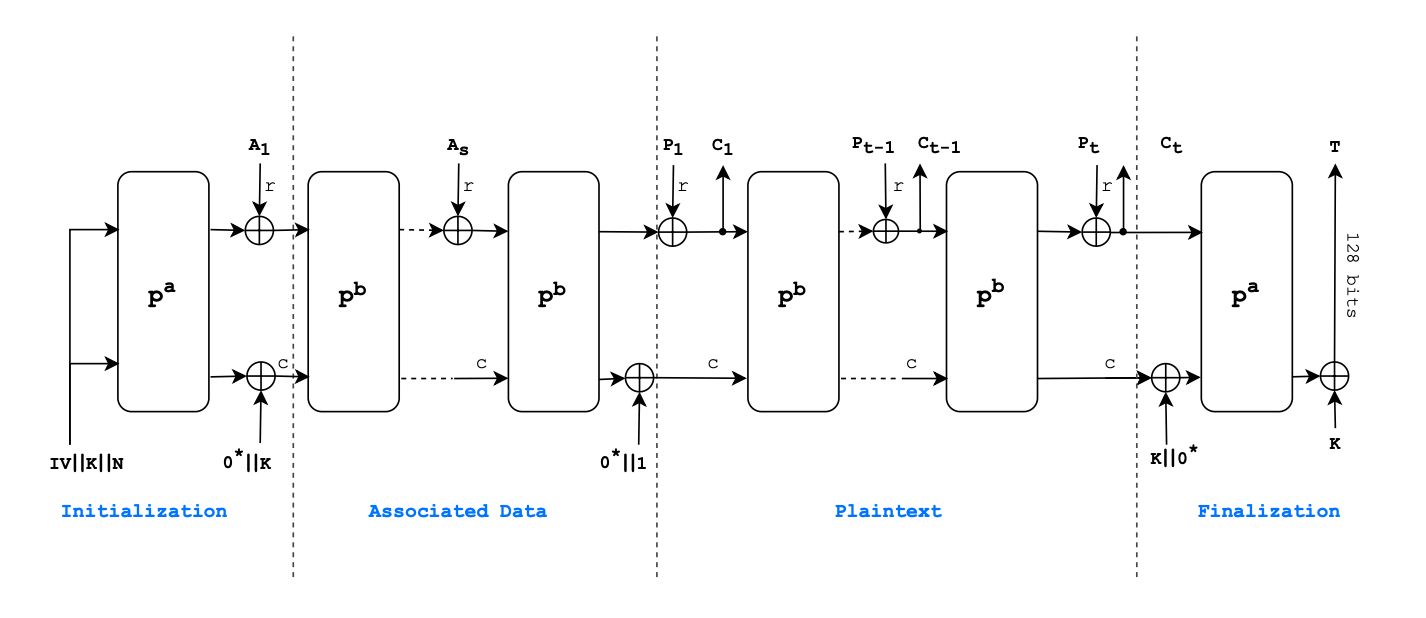}
       \caption{Encryption Process of ASCON}
\end{figure}
pa and pb are permutation blocks of size rate (r) + capacity (c) and have no.
of rounds as a and b respectively. The initialization takes as input a Initialization
Vector IV secret key K and the public nonce N. The initialization ensures that we
start with a random-looking state at the beginning of the initialization phase.
Next in processing of Associated data, the cipher absorbs all the associated data
r-bytes at a time. In the encryption phase, plaintext block Pi of size r-bits is
XORed with the first r-bits output of the permutation pb and ciphertext is
obtained. At last comes the finalization phase where the key is XORed with
the last c-bits of the output to obtain the tag which is used for integrity check.
Since we are focused only on the encryption of PDF we ignore the tag in our
experiments.

\subsection{XOODYAK}\label{subsec2}
 XOODYAK is one of the finalists of LWC competition held by NIST. It is based on
a combination of sponge structure and XOODOO permutation. In 2018 Daemenet. al introduced XOODOO \cite{xoodoo}, inspired by Keccak \cite{keccak} and Gimli \cite{gimli} which is a 48-byte permutation specially designed for cryptographic purposes. It is
based on Cyclist operation mode. The state on which XOODOO operates is a
3-dimensional array of dimensions 4 × 3 × 32. XOODYAK consists of 12 rounds
and works on hash and keyed modes. The encryption process of XOODYAK is
same as that of ASCON i.e. duplex construction where it has rate of 16 bytes
and capacity of 32 bytes. The total permutation happens on 384 bits .
 
\section{The Cipher Switch Strategy}\label{sec3}
For replacing AES with ASCON and XOODYAK we needed to unroll the whole encryption process. We needed to do so to pinpoint the part of the code where actual encryption happens after Encryption Key has been generated and a particular string or stream is retrieved from the PDF. One very interesting part of PDF encryption is that it uses different keys to encrypt different strings and streams. This is done by appending the initial Encryption Key generated with low-order 3 bytes of the object number and the low-order 2 bytes of the generation number in that order, low-order byte first. This extended key is then used to encrypt the respective string or stream. After the ciphertext is obtained it is checked whether the plaintext had some type of encoding (which in most streams are encoded with flate encoding). For this we need to update the compression filter at the time of replacing the original stream with it’s respective ciphertext.
This step must be taken into consideration since, while replacing the plaintext
with the ciphertext in the PDF, encoding happens again which appends extra
bytes at the beginning and at the end of the stream. This changes the overall
ciphertext structure and on decryption we get different result that the required
plaintext. Due to this PDF readers are not able to render the PDF. Since in
most cases the streams are encoded beforehand there is no need for decoding
them. The encoded stream acts as plaintext. On opening the encrypted PDF
with correct key the streams are first decrypted, then decoded and lastly the
PDF is rendered by the reader. If there arises the need to encode again, one
needs to mention the respected algorithm after the crypt value in the Filter
entry of the stream dictionary. It may be the case that all the allowed strings
and streams are encrypted, in which case it is not necessary to mention crypt
in the Filter entry. It becomes necessary when only some of the allowed strings
and streams are encrypted. A brief overview of The Cipher Switch strategy can
be summarized as follows:

\begin{enumerate}
    \item Generation of O value from provided owner password
    \item Generation of Encryption Key
    \item Generation of U value from provided user password
    \item Retrieve a string or stream from the PDF
    \item Update the Encryption Key
    \item Select the required cipher to encrypt
    \item Obtain the ciphertext
    \item Since we are benchmarking the performance, we also need to consider the time for decryption. Decryption is done and the plaintext obtained is discarded.
    \item Replace the plaintext with the obtained ciphertext in the PDF
    \item Add Encrypt dictionary as a separate object with all the required attributes in the PDF
    \item Add the reference to the Encrypt dictionary to the trailer dictionary
\end{enumerate}
\vspace{0.2cm}
 On decryption we do not retrieve encrypted streams from the PDF since it
will just add overhead and would contribute nothing to the performance. This is
due to the fact that retrieving strings or streams is a constant part of our work
and since time taken for this will be same regardless of the cipher used. Thus
we just decrypt the obtained ciphertext and consider only the decryption time.

\vspace{0.5cm}

\vspace{0.5cm}


\section{Performance Benchmarking of ASCON and
XOODYAK with AES}\label{sec4}
Since iText already has AES implemented, we only needed JAVA implementation
of ASCON and XOODYAK. On quering over google we found implementations of
ASCON and XOODYAK on github. We chose specific classes and modified bits
and parts of their code to suit our needs and also got rid of any statements which
uses I/O such as print statements. We specifically chose these 2 ciphers since
ASCON has been announced as LWCstandard and specification of XOODYAK
is similar to SHA3 which is widely used for hashing purposes. For ASCON we
use \cite{asconjava} and for XOODYAK we use \cite{xoodyakjava}. For the rest of the implementation we have
used iText library for PDF manipulation, mainly encryption.

\subsection{Benchmarking Tools}\label{subsec2}
For benchmarking performances with each algorithm running seperately we used
Java Microbenchmark Harness (JMH). JMH allows different types of benchmark-
ing modes such as Average time, Single Shot time, Throughput and Sampling
time. For profiling our code we use YourKit Java Profiler 2022.9-b182 integrated
with eclipse. Since we cannot profile while running JMH, we created a different
version of the code to profile each algorithm independently. \vspace{0.2cm}

\subsubsection{Java Microbenchmark Harness}\label{subsubsec2}
JMH is a tool which is used to benchmark
pieces of JAVA code correctly. It is developed by the same people who implement
JAVA Virtual Machine (JVM) \cite{jmh}. Since we use Maven to set up a project we
needed to add dependencies in pom.xml. During benchmarking we found that
we do not use the plaintext retrieved after decryption of the ciphertext but we
need the decryption process to happen for our benchmarking. This plaintext
obtained will be discarded by the compiler as dead code elimination. But JMH
provides Blackhole object which consumes the not-required plaintext. By default
JMH has Blackhole mode on auto detect. JMH does some warmup iterations of
the code to be benchmarked, the quantity of which can be defined by the user.
For our use we have set warmup iterations to 5 times. After warmup actual
measurement begins. We can define the no. of times to iterate the code. Results
of these iterations is taken into consideration for the final result. \vspace{0.2cm}

\subsubsection{YourKit Java Profiler}\label{subsubsec2}
YourKit Java Profiler provides a wide range of options to
profile JAVA applications either locally or remotely. Among various options it has
Telemetry and Performance graph, Database queries and web requests, memory
and CPU profiling, exception profiling etc \cite{yourkit}. Moreover it can be integrated with
IDE to provide ease of use. The results are saved as snapshots. The main reason
for us to incline towards this profiler is that it provides call tree of the program
flow with each method showing time duration from beginning of the execution
of the method till it completes executing. To get more reliable results we profiled
the code where each cipher is executed separately. We found the results to be
much organized and easy to understand.

\subsection{Comparative Analysis}\label{subsec2}
The systems used for obtaining the following results are .
\begin{enumerate}
    \item Intel(R) Core(TM) i5-7200U CPU @ 2.50 GHz, 2.71 GHz with 8 GB of RAM
    \item AMD Ryzen 7 3700U with Radeon Vega Mobile Gfx @ 2.30 GHz with 16 GB of RAM
    \item Intel(R) Core(TM) i7-5500U CPU @ 2.40 GHz, 2.40 GHz with 16 GB of RAM
\end{enumerate}
The benchmarked function retrieves each string and stream individually and
passes them as argument to another method where encryption and decryption
happens. The ciphertext thus obtained is stored in a byte array. This ciphertext is
decrypted and is discarded. The ciphertext is then passed to the parent function.
After the flow returns to the parent function we replace the original plaintext
with the ciphertext and add required PDF objects for the reader to interpret
properly. Since PDF standard does not allow any other algorithm than AES,
our encrypted PDF using algorithms other than AES won’t open in any PDF
readers existing till date. We ignore the retrieving of streams and strings while
decryption because it is independent of the algorithm used and it will be same
for every algorithm. We measure only the decryption time of the ciphertext

\subsubsection{YourKit Java Profiler results}\label{subsubsec2}

\textbf{JMH results} We have measured for Average Time taken in Fig. 10 and Single
Shot Time in Fig. 11 We have specifically taken these modes since they depict
encryption (and decryption) of multiple and single PDF respectively. The results are presented with error margin. The graphs are plotted without the error margin. The unit for each result is \textmu s/op where op is known as operation. Each operation represents one iteration of the implementation.

\begin{figure}
  \centering
      \includegraphics[width=0.99\textwidth]{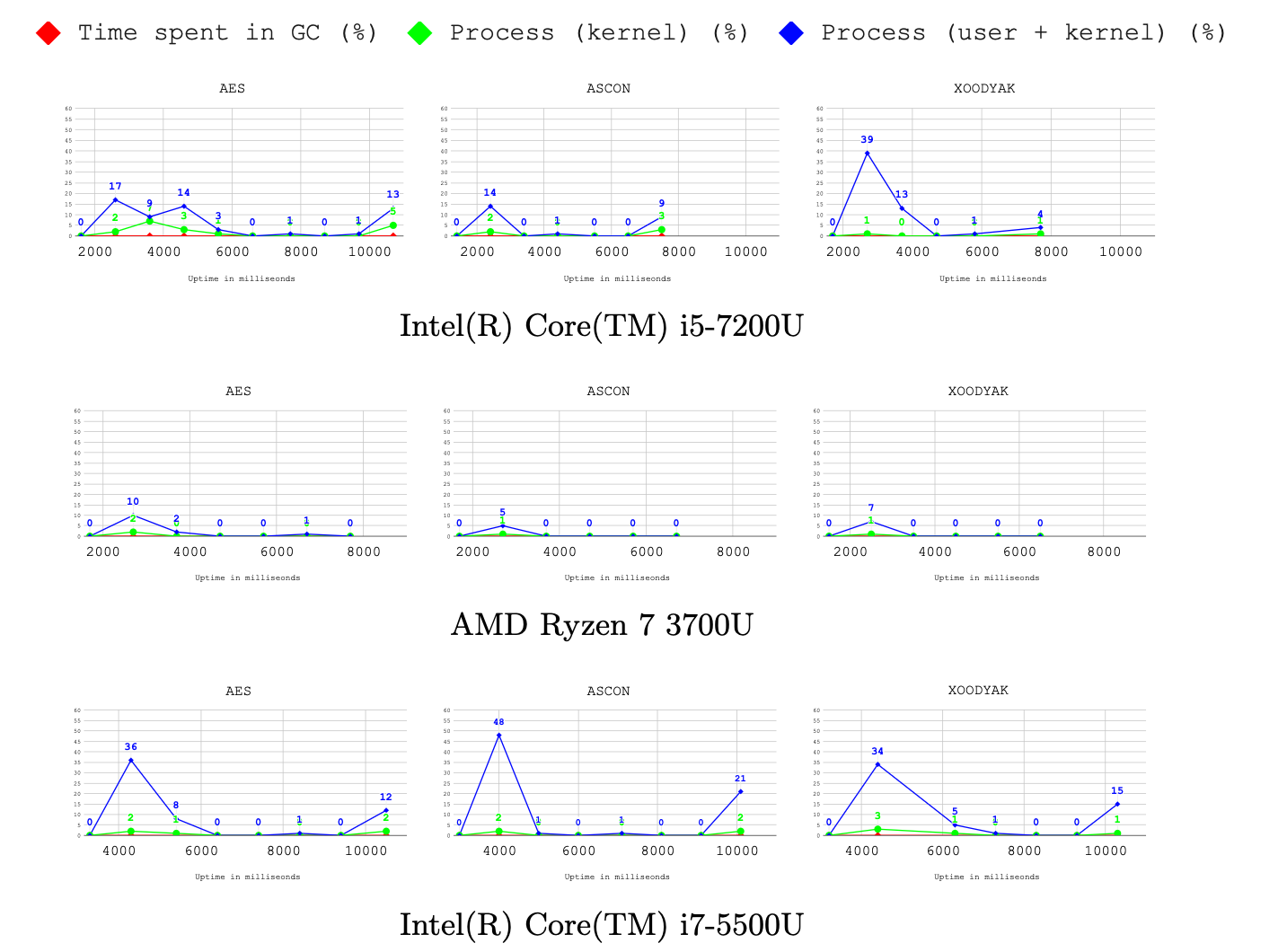}
       \caption{\textbf{Encryption Time Comparison of AES, ASCON, and XOODYAK}}
\end{figure}

\section{Discussion}\label{sec12}

In YourKit profiler results, there is a certain spike in the beginning of the graph
which depicts the amount of usage of CPU which the process and process +
kernel puts on the CPU. On observing carefully we can deduce that using AES
to encrypt and decrypt PDF, the process consumes more CPU for a longer period
of time when compared with ASCON and XOODYAK. On comparing the graphs
we can see that on using AES, the process takes longer to complete than ASCON
and XOODYAK. This difference is very small in AMD Ryzen 7 3700U but can
be seen clearly in Intel(R) Core(TM) i5-7200U.
12
In JMH results we see that on both the modes (i.e. Average time and Single
Shot time) ASCON and XOODYAK had taken less time to execute and measure
according to our measurement parameters. For Average time each algorithm is
run 10 times over a same PDF. Each algorithm is forked twice which gets us a
total iteration of 20 times in two instances. Similarly for Single Shot time each
algorithm is run 10 times in 2 iterations from which results are obtained.
Lastly we can infer that using ASCON as PDF encrption and decryption
algorithm we can have best performance overall. Comparing the results we see
that in every one of the cases AES performs poorly with respect to ASCON and
XOODYAK. Using ASCON will enable faster encryption and decryption of PDF
file format. Since decryption is used every time a PDF is opened this becomes
an important part for making PDF more lightweight
\begin{figure}
  \centering
      \includegraphics[width=0.99\textwidth]{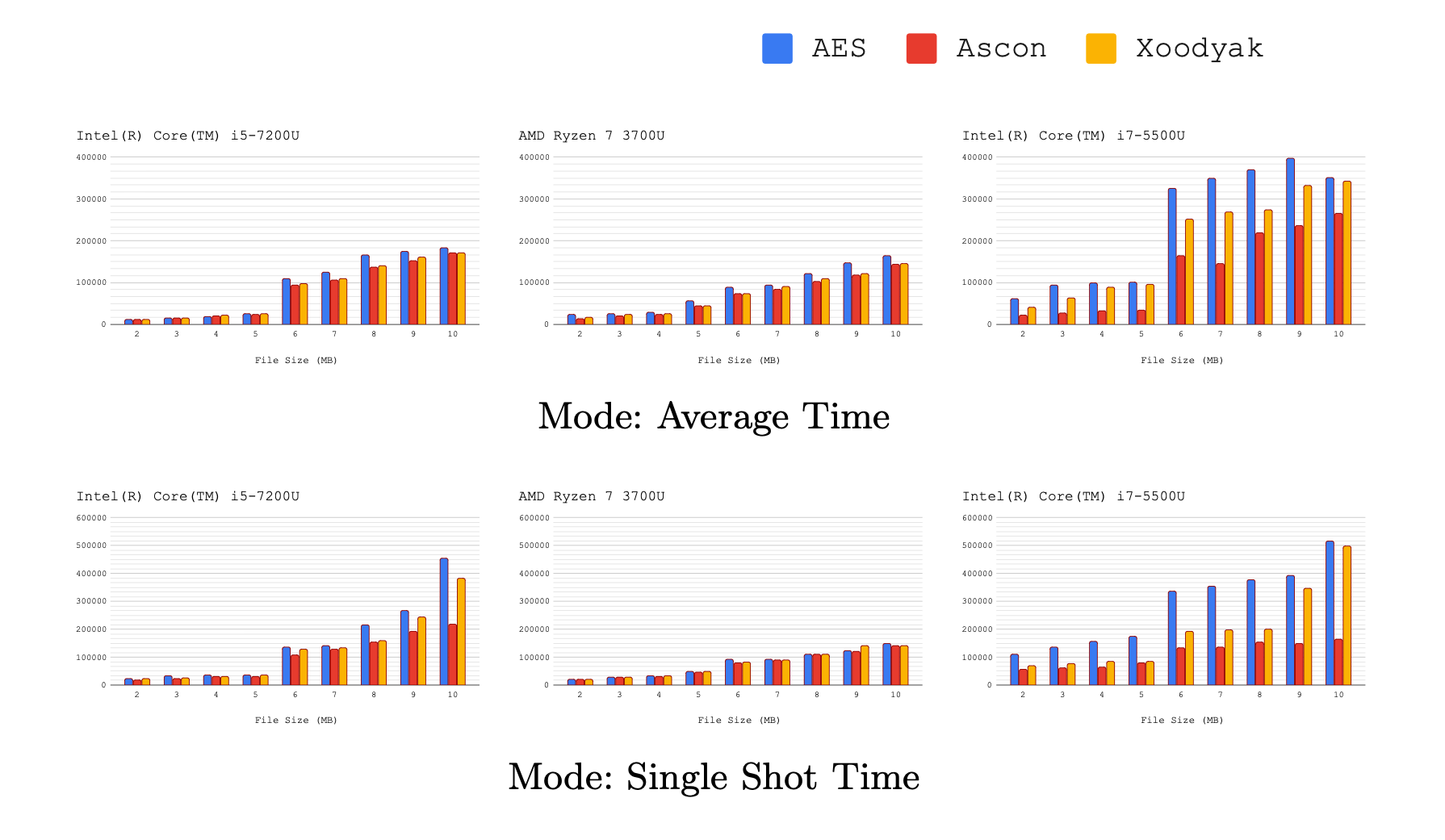}
       \caption{\textbf{Impact of AES, ASCON, and XOODYAK on PDF File Size (in MB})}
\end{figure}

\newpage
\section{Future Scope}\label{sec13}
PDF is widely used for contracts, official documents, and forms, often containing sensitive information that requires strong security measures. As security needs evolve, we have introduced new ciphers like ASCON (now an LWC standard) to improve PDF encryption efficiency. While the flexible structure of PDFs makes encryption and decryption complex, our approach focuses on maintaining security without compromising the lightweight nature of PDFs. Our work demonstrates that using ASCON for PDF encryption results in faster and more efficient processing. There is potential to explore implementation of hybrid encryption schemes that combine the strengths of both symmetric and asymmetric cryptography. This would aim to balance the trade-offs between encryption speed, security, and computational resources more effectively.

Additionally, there is an opportunity to extend the benchmarking to include real-time performance metrics under different network conditions, particularly in cloud-based PDF processing systems. Finally, as new versions of PDF standards are developed, it would be valuable to continuously adapt and test these cryptographic methods to ensure compatibility and to leverage any advancements in encryption technologies.\\
\begin{figure}\label{tab:average-time}
  \centering
      \includegraphics[width=1\textwidth]{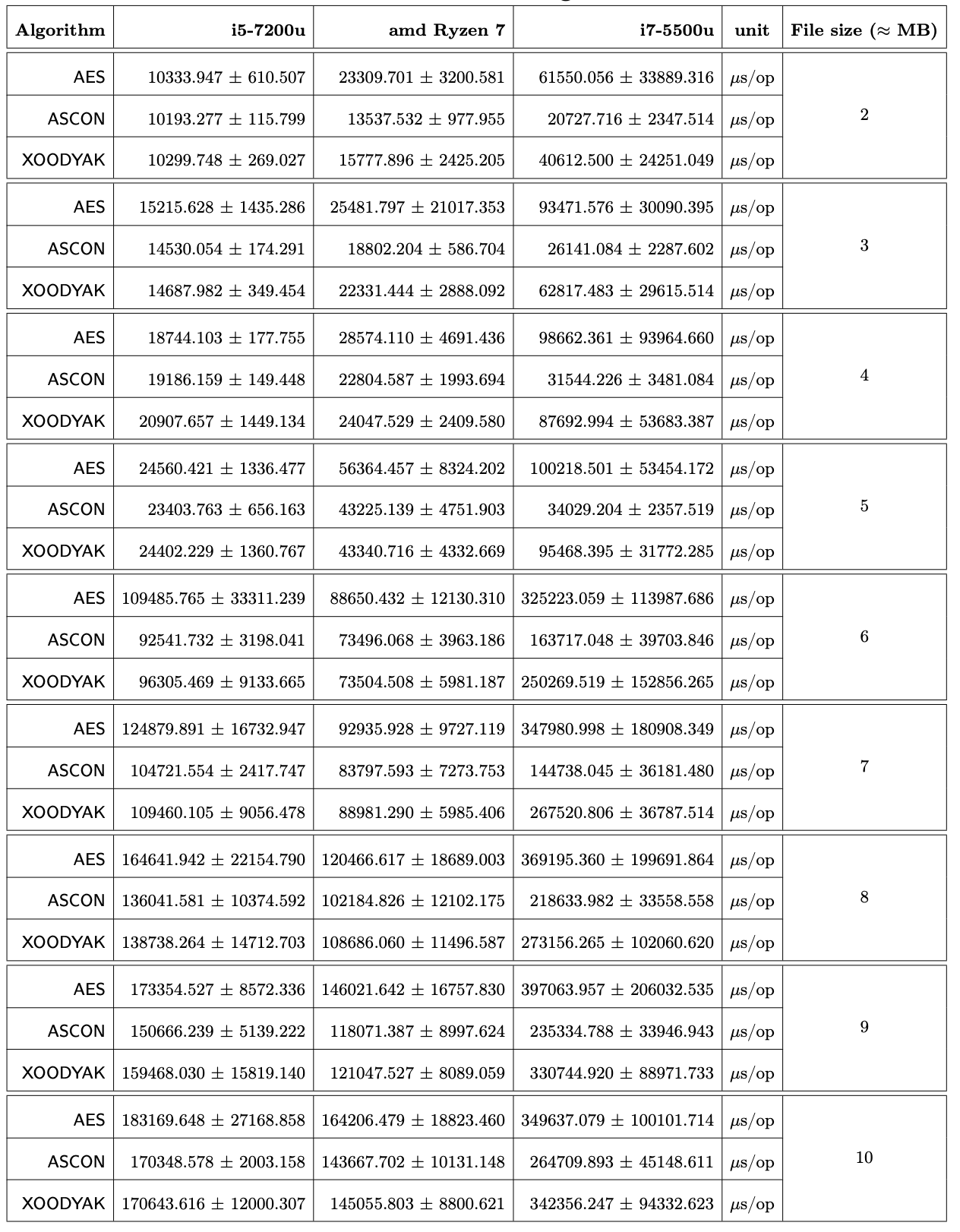}
       \caption{\textbf{Mode: Average Time}}
\end{figure}

\begin{figure}
\label{tab:single-shot-time}
  \centering
      \includegraphics[width=1\textwidth]{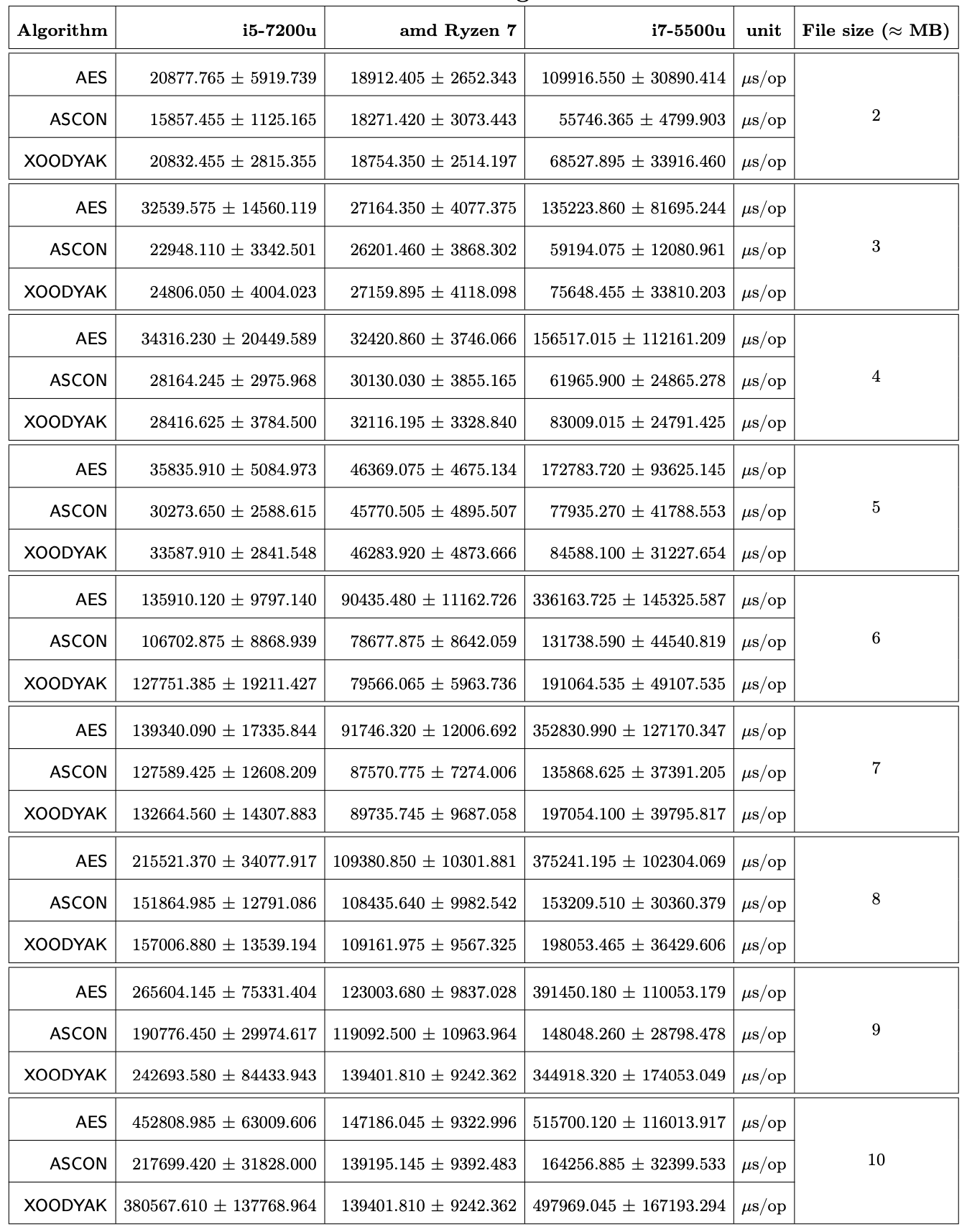}
    \caption{{\textbf{Mode: Single Shot Time}}}  
    
\end{figure}

\backmatter



\bibliography{sn-bibliography}

\end{document}